\documentclass[conference]{IEEEtran}
\usepackage{cite}

\ifCLASSINFOpdf
  \usepackage[pdftex]{graphicx}
   \graphicspath{{C:/Users/Administrator/Desktop/latex/eps}}
  \DeclareGraphicsExtensions{.pdf,.jpeg,.png,.eps}
\else
\fi
\usepackage{CJK}
\usepackage[numbers,sort&compress]{natbib}
\usepackage{cite}
\usepackage{amsmath,amssymb,enumerate}
\usepackage{cite,graphics,graphicx}
\usepackage{graphicx}
\usepackage{subfigure}
\usepackage{listings}
\usepackage{makeidx}
\usepackage{hyperref}
\usepackage{algorithm}
\usepackage{algorithmic}
\usepackage{float}
\usepackage{multirow}
\usepackage{amsmath}
\usepackage{amssymb}
\usepackage{xcolor}
\usepackage{epsfig}
\usepackage{graphics}
\usepackage{listings}
\usepackage{algorithm}
\usepackage{algorithmic}
\hypersetup{linkcolor=black, %
            citecolor=black, %
             pdftitle={Title}, %
            pdfauthor={Name}, %
           pdfsubject={Subject}, %
          pdfkeywords={Key words}}
\begin{document}

%
\title { Hybrid Ant Colony Algorithm Clonal Selection in the Application of the Cloud's Resource Scheduling}

\author{\IEEEauthorblockN{Jianbiao Lin{$^{1}$},
Yukun Zhong{$^{2}$},
Xiaowei Lin{$^{3}$},
Hui Lin{$^{4}$},Qiang Zeng{$^{4}$}
}

\IEEEauthorblockA{
Computer Science and Engineering Department\\ Sichuan University Jinjiang College, Penshan 620860, China\\
3. College of Economics and Management\\ Dalian JiaoTong University ,Dalian 116028, China\\
Email: zeroyuebai@hotmail.com}
}


%


\maketitle

\begin{abstract}
 In this paper, thinking over characteristics of ant colony optimization Algorithm, taking into account the characteristics of cloud computing, combined with clonal selection algorithm (CSA) global optimum advantage of the convergence of the clonal selection algorithm (CSA) into every ACO iteration, speeding up the convergence rate, and the introduction of reverse mutation strategy, ant colony optimization algorithm avoids local optimum. Depth study of the cloud environment ant colony clonal selection algorithm resource scheduling policy, clonal selection algorithm converges to solve optimization problems when sufficient condition for global optimal solution based on clonal selection algorithm for various applications such as BCA and CLONALG algorithm, using these sufficient condition to meet and simulation platform CloudSim achieve a simulation by extending the cloud. Experimental results show that this task can be shortened fusion algorithm running time cloud environment, improve resource utilization. Demonstrate the effectiveness of the method.\\\\
 keywords- Ant Colony Optimization, Clonal selection algorithm, Cloud Environment, Resource Scheduling

\end{abstract}


\section{Introduction}
Cloud computing is parallel computing , Distributed Computing  and Grid development, or that these commercial implementations of computer science concepts. Cloud computing is virtualization , utility computing , IaaS (Infrastructure as a Service), PaaS (Platform as a Service), SaaS (Software as a Service) concepts such as evolution and jumped in mixed results. Cloud computing is a model of business computing, distributed computing tasks will constitute a large number of computer resource pool, so that the system can obtain a variety of applications as needed computing power, storage space and a variety of software services.

Cloud computing emphasizes shared, heterogeneous, dynamic collaboration of resources, which bring convenience to users, but also for resource scheduling technology put forward higher requirements. Cloud computing refers to the resource scheduling of a particular cloud environment, and the use of resources in accordance with certain rules, between different resources in the resource and the user performs which needs the adjustment process. Most current resource scheduling policy is a combination of virtual machines by scheduling techniques on a certain level scheduling policy to try to do for the virtual machine resource scheduling internal applications \cite{foster2008cloud}, a general lack of fine certainty.

Ant colony optimization algorithm is an adaptive search algorithm based groups,the community food is ants acquisition process simulation. Research shows that ant colony optimization algorithm has a strong ability to find better solutions of combinatorial optimization problems in solving, but also has a distributed computing, easy combination with other methods, robustness, etc., in a dynamic environment exhibit a high degree of flexibility and robustness, the successful resolution of a number of combinatorial optimization problems. And the nature of cloud computing resource scheduling problem is selected from the resources assigned to tasks in a variety of combinations of relatively good performance of a dynamic combination, to solve the problem from the perspective of the ant colony optimization algorithm is very suitable for solving a cloud environment the resource scheduling problem \cite{wang2007study} But the ant colony algorithm in solving large-scale combinatorial optimization problems easy to fall into local optimum search is slow and defects, and therefore the basis of ant colony optimization algorithm, the introduction of clonal selection algorithm and part of path reversal of variation. Complementary advantages of the two algorithms that solve optimization ability and speed greatly improved, improved scheduling efficiency \cite{dutta2011genetic}.

Clonal selection algorithm \cite{de2002learning} (CSA) is an artificial immune system artificial immune system, AIS) is an important algorithm, has been widely used in AIS and achieved some success (such as pattern recognition and function optimization and other fields) . It is different biological mechanisms and evolutionary computation depends compare clonal selection algorithm, but has shown a lot of useful features, such as better maintain the diversity of the population, which can effectively overcome such as prematurity and other evolutionary computing itself is difficult to solve issue.

So far, the analysis and research work on the theoretical aspects of clonal selection algorithm is still small, and only literature \cite{villalobos2004convergence} for a specific immune systems multi-objective optimization algorithm (MISA),which the use of Markov chain gives its full convergence proof. \cite{clark2005markov}For a clonal selection algorithm ---- B cells in the form of a simplified algorithm (BCA),and describes a mathematical model BCA algorithm ---- Markov chain model, and proposes the use of an algorithm change in BCA super mutation operator ---- contiguous area hypermutation operator (CRHO), and the establishment of a state transition matrix BCA algorithm using a new method, and the use of absolute convergence BCA algorithm can be used as like \cite{de2002ant} that is current optimal solution to maintain a separate set of distinguished persons memory.

\section{Standard Ant Colony Optimization Algorithm And General Clonal Selection Algorithm}\label{SEC: Standard Ant Colony Optimization Algorithm And General Clonal Selection Algorithm}

Ant Colony Optimization (ACO) is by simulating the natural foraging behavior of real ants, put forward a novel simulated evolutionary algorithms, swarm intelligence algorithm is a major theoretical research areas. TSP problem with this method, allocation, job-shop scheduling problem, and achieved good results. Although the study is not long, but now research shows that ant colony algorithm in solving complex optimization problems (especially discrete optimization problems) has certain advantages, indicating that it is a promising algorithm.

The clonal selection also called Strain drop selection, and the immune cells can be randomly generated clone diversity Cloning and expression of each immune cells for a particular antigen-specific receptor. For a particular antigen, and expression of specific immune cell receptors which specifically bind, resulting in such a large number of immune cells are activated and amplified. Activation of the different antigens of different immune cell clones. Immune tolerance is due to the occurrence of a premature immune cell clones that bind to the antigen with its own immune cells before cell stage amplification occurred abortion. Clonal selection algorithm (CSA) to fully meet the convergence conditions globally optimal solution is a can converge in a limited generation mature algorithms.

Here we focus on these two key issues discussed in detail.

\subsection{Independent analysis algorithm ACO}\label{SSEC: Independent analysis algorithm ACO}

Ant Colony by positive feedback mechanisms and collective autocatalytic behavior pheromones to find the optimal path. M. Dorigo and other scholars have put forth a simulated ant colony foraging behavior of this AS (ant system) algorithm and ACS algorithm, and two custom algorithms for ACO \cite{casanova2000heuristics} algorithms.

Through ACS system model with a plane as an example to solve the plane n cities TSP, first introduced symbolic amount of:

m: indicates the number of the colony of ants; $ d_{ij} $, ($ ij = n, 1, 2, \ldots $): indicates the distance $ i $, $ j $ between cities;

$ \tau_{ij}(t) $: represents the residual amount of information in the city at time t i, j wiring, general admission $ \tau_{ij}(0) = C $ (C is constant);

$ \eta_{ij} $: at time $ t $ ant transferred from city $ i $ to city $ j $ heuristic information, generally take $ \eta_{ij} = 1/d_{ij} $;

$ p_{ij}^k(t) $( $ t $) : $ t $ represents the time the ants transfer from city $ i $ to city $ j $, the probability is calculated as

$\alpha $, $\beta $ represent the importance of the importance of information and inspiration residual information.

Step 1. Initialization. Initialize the pheromone on each edge of a small constant value, the m ants randomly placed into n cities, and the starting point of the city is set to tabu $ _{k} $ in taboo list.

Step 2. Ant according to equation (1) to select the next city, and modify the taboo list.

Step 3. After each ant has completed an edge, according to equation (2) update the edge pheromone.
\begin{equation} \tau_{ij}(t + 1) = (1 - \rho )\tau_{ij}(t) + \rho \Delta \tau_{ij}^k \label{eq:2}\end{equation}
$$ \Delta \tau_{ij}^k = \begin{cases}
Q/l_{jb},& \mbox{When the ant $ k $ through the $ ij $ city} \\
0,&  \mbox{Otherwise}
\end{cases} $$

$ l_{jb} $ city where is the ant k from the beginning to the current city has gone through a path length.

$\rho $ represents persistent information.

Step 4. Calculate the best path. When all the ants finish all the cities, according to equation (3) to calculate the optimal path length and retention.
\begin{equation} l_{min} = \min l_{k}, k = \{1, 2, \ldots, m \}, \label{eq:3}\end{equation}

$ l_{k} $ representatives ant $ k $ are taking the path length.

Step 5. When all the ants completed after all of the city, only the optimal path on the pheromone according to the formula (4) is updated.
\begin{equation} \Delta \tau_{ij}{new} = (1 -\alpha)\tau_{ij}^{old} + \alpha\Delta \tau_{ij}^k \label{eq:4}\end{equation}
$$ \Delta \tau_{ij}^k = \begin{cases}
1/l_{k},& \mbox{If $ ij\in $ global optimal path} \\
0,& \mbox{Otherwise}
\end{cases} $$

$ \alpha $ is a global pheromone evaporation coefficient, $ l_{k} $ optimal path length.

Step 6. If the number of iterations set unfinished, then empty the taboo list, repeat the process.

\subsection{The ant colony and clonal selection fusion algorithm in task scheduling and cloud environments}\label{SSEC: The ant colony and clonal selection fusion algorithm in task scheduling and cloud environments}

ACO has many advantages. Currently, ant colony optimization algorithm proposed improved variety, achieved very good results. However, due to the randomness algorithm great in solving large-scale optimization problems easily be limited to local optima, slow convergence defects. This paper introduces clonal selection algorithm, combined with its fast random global search capability,and the clonal selection algorithm into every iteration of the ant colony optimization algorithm, which improving the convergence rate, and ensure the accuracy of the original algorithm to solve. While the introduction of reverse mutation strategy \cite{li2003combination}, and to avoid the risk of the optimal algorithm into local, what is maintain and increase the diversity of the population.

Fusion clonal selection algorithm using ant colony to search for the best current allocation policy strategy, at the same time, factors affecting the resource state can always be described by the information, then the scheduler can simply, quickly get predictable results. Due to the unique nature of the cloud environment, the need for TSP problems associated with cloud characteristics between environmental resource scheduling problems for improved ant colony optimization algorithm \cite{casanova2000heuristics}. These differences are mainly in the following three aspects:

(1) In the TSP, there are edges and these edges are connected up and have different distances between cities, and in a cloud environment, the resources are not fixed network topology. Therefore, in a cloud environment to take advantage of resources activities to simulate the interaction topology.

(2) In the TSP problem, the pheromone represent between cities with the distance between cities , and in a cloud environment need to use computing and communications capability to represent the correlation coefficient resources pheromone.

(3) Inspired information also plays a very important role,which in a cloud environment, the need to resource the inherent properties (such as computing and communications capabilities, etc), and how to represent the heuristic information in the ant colony algorithm.

Based on the above three points, ants based on current information resources, decided to allocate to the calculation method of the cloud computing environment, the next task probability of existing resources also make corresponding changes are as follows:
$$ p_{jk}^i = \begin{cases}
\frac{\left[\tau_{j}(t)\right]^\alpha \left[\eta_{j}\right]^\beta}{\sum\limits_k \left[\tau_k(t)\right]^\alpha \left[\eta_k\right]^\beta},& \\
\qquad \mbox{$ j $, $ k $ $\in $ cloud computing resources available} & \\
0,\quad \mbox{Otherwise}
\end{cases} $$

$ \tau j(t) $ is the pheromone concentration of resources at time t;

$ \eta j $ represents the intrinsic properties of the resource;

$ \eta j = \tau j(0) $;

$ \alpha  $ indicates the importance of pheromones;

$ \beta  $ indicates the importance of the inherent properties of the resource.

In this paper, the use of the general method of clonal selection ant colony algorithm to solve the task scheduling cloud computing. such as in the literature \cite{de2002learning}, when the CLONALG used for pattern recognition, we should remain independent of memory cells and antigen populations, but will the need for multi-mode function optimization CLONALG these two groups; Again, a special hypermutation used in BCA \cite{kelsey2003immune} the operator --- continuous area hypermutation operator (CRHO). In this paper, the general behavior of each generation of clonal selection algorithm. The description of a general, encompasses all of the CSA operator and its changing operator\cite{}. Its clonal selection algorithm behavior is as follows:

Clonal Selection Algorithm(T$ _{\beta} $)//T$ _{\beta} $ is the algorithm parameters

t:=0;

$ P^{(t)} $: = Initialize Pop();

While( - Termination Condition())  do

\advance\leftskip1em Evaluate(P$ ^{(t)} $);

$ P^{(clo)} $: = Cloning(P$ ^{(t)} $);

$ P^{(hyp)} $: =Hypermutation(P$ ^{(clo)} $);

Evaluate(P$ ^{(hyp)} $);

(P$ _{a} $$ ^{(t)} $,P$ _{a} $$ ^{(hyp)} $): =Aging(P$ ^{(t)} $,P$ ^{(hyp)} $,T$ _{\beta} $);

$ P^{(t+1)} $; = Selection(P$ _{a} $$ ^{(t)} $,P$ _{a} $$ ^{(hyp)} $);

t = t+1;

\advance\leftskip-1em End while

Cloning operator generates only a copy of the individual, which the individual does not change any of the values, so it will not lose the optimal solution.

Hypermutation operator only operator in the middle of cloning generated. Group p $ ^{(clo)} $ to perform the operation, nor alter the individual by any other operator (including its own) generated.

Aging operator removed the old individuals despite, but because of the optimal candidate solutions age population of each generation is set to 0. It is impossible to lose the optimal solution.

As for the selection operator, because it removed the minimum affinity individuals, replaced by a new generation of individuals randomly,  the optimal solution is not lost, even with no redundancy selection operator. If there are multiple optimal solutions group, then at least there is an optimal solution is also survived.

\section{Simulation results}\label{SEC: Simulation results}

\begin{figure}[!htb]
    \centering
    \includegraphics[width=.9\linewidth]{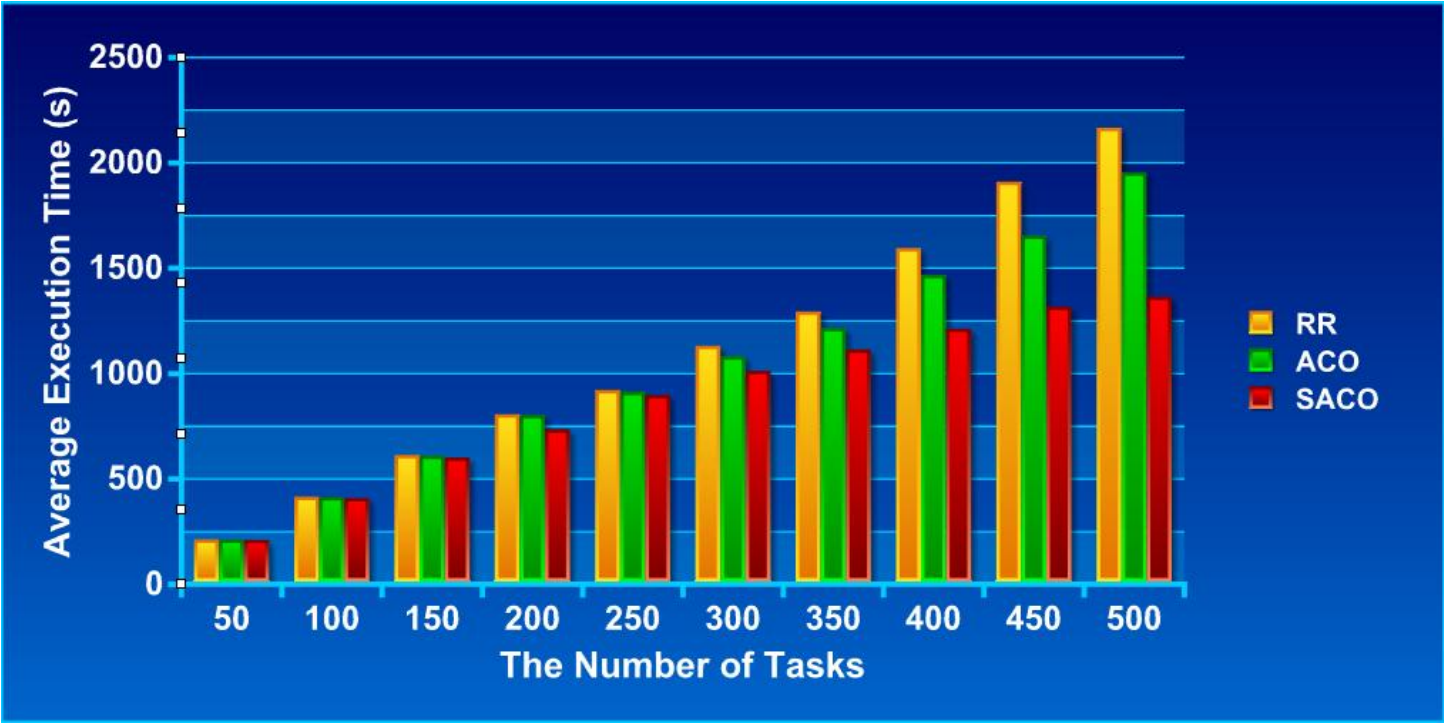}\\
    \caption{\label{fig:2-1}Comparison of the average execution time tasks}
\end{figure}

Cloudsim by the University of Melbourne and Gridbus jointly launched cloud simulation software platform, the latest version Cloudsim3.0. This experiment on the use of this platform to simulate experiments. By rewriting DataCenterBroker, Cloudlet, VmScheduler and other classes to achieve different task scheduling algorithm, ant colony clonal selection of the proposed hybrid algorithm in cloud computing resource scheduling simulation tests to validate the superiority of this algorithm.In this experiments were used CloudSim existing Round-Robin(RR) scheduling algorithm, ACO algorithm and MACO task scheduling algorithm. The average execution time of each task in \figurename{} \ref{fig:2-1}. Experimental environment for the Windows 7 operating system, processor Intel Core i7-4700M, 2.40GHz, memory is 8GB.

As can be seen from the histogram in \figurename{} \ref{fig:2-1}, RR algorithm with the number of tasks increases, the time it takes the more. The ACO algorithm, because at the beginning of fewer pheromones, slow task execution late with the increasing pheromone enhanced positive feedback, the time rate of increase is less than the RR algorithm. SACO improved algorithm is clearly better than the first and two algorithm execution time is shorter, higher efficiency.

\section{Acknowledgment}\label{SEC: Acknowledgment}

These research subject was supported by Sichuan University Jinjiang College, the department of Computer Science$\&$Engineering. Thanks for Prof.Bingfa Lee¡¯s suggestions and guidance , and Ruilian Han $\&$ Yonggui Wang and Xianjin Fang $\&$ Longshu Lee whose books what give me a lot of inspiration.





%

\bibliographystyle{IEEEtran}

\bibliography{BB}


%

\end{document}